\documentclass[12pt]{article}
\usepackage{amsmath,amsfonts, amssymb,graphicx,geometry,authblk,color,soul,comment,hyperref} 
\usepackage{bm}

\begin{document}

\title{Softening of the Hertz indentation contact in nematic elastomers}  
\author{Ameneh Maghsoodi$^{1}$, Mohand O. Saed$^{2}$, Eugene M. Terentjev$^{2}$, and Kaushik Bhattacharya$^{3 \ *}$}
\affil{$^1$Department of Aerospace and Mechanical Engineering, University of Southern California, Los Angeles, CA 90089, USA}
\affil{$^2$Cavendish Laboratory, University of Cambridge, Cambridge CB3 0HE, U.K.}
\affil{$^3$Division of Engineering and Applied Science, California Institute of Technology, Pasadena, CA 91125, USA}
\date{\today}

\maketitle

\begin{abstract}
\noindent Polydomain liquid crystalline (nematic) elastomers have highly unusual mechanical properties, dominated by the dramatically non-linear stress-strain response that reflects stress-induced evolution of domain patterns. Here, we study the classical Hertz indentation problem in such a material.  Experimentally, we find that polydomain nematic elastomers display a smaller exponent than the classical 3/2 in the load vs. indentation depth response.  This is puzzling: asymptotically a softer stress-strain response requires a larger exponent at small loads. We resolve this by theory where three regimes are identified -- an  initial elastic regime for shallow indentation that is obscured in experiment, an intermediate regime where local domain pattern evolution leads to a smaller scaling in agreement with experiments, and a final stiffening regime where the completion of local domain evolution returns the response to elastic.  This three-regime structure is universal, but the intermediate exponent is not.  We discuss how our work reveals a new mechanism of enhanced adhesion for pressure-sensitive adhesion of nematic elastomers.
\end{abstract}

The elastic softness of nematic elastomers is a remarkable phenomenon, with very few analogies in any other material \cite{soft1994,WTbook}. It originates from the unique feature of the network microstructure, where the base elastic element -- the `average' chain between its terminal crosslinks, is uniaxially anisotropic in the nematic liquid crystal phase. A macroscopic deformation can accrue when such chains rotate the axis of their anisotropy, but no entropic elasticity arises if the average shape of the rotating chain is preserved. This phenomenon has been extensively studied over the years, both experimentally and theoretically.   In particular, many nematic elastomers display `semi-softness': the non-ideal response when the symmetry of average chain rotation and the associated shear deformation is preserved, but the presence of internal constraints produces an initial linear elastic regime, a threshold strain, and a non-zero stress plateau, until the full chain re-alignment could be reached. Many microscopic mechanisms for this non-ideal `semi-softness' have been identified, ranging from the polydispersity of network chains \cite{Verwey1997a} to the effect of anisotropic crosslinkers \cite{Verwey1997b}, to the entanglement of nematic chains \cite{Kutter2001}.

A separate, equally unique feature of liquid crystal elastomers (LCEs), is the fact that when no particular anisotropic field is employed during their network formation, the resulting phase becomes an equilibrium polydomain. This is fundamentally different from an initially disordered Schlieren texture of a liquid nematic, which is a non-equilibrium kinetic vestige of nucleation during the first order transition from the isotropic phase. In nematic elastomers, the polydomain texture is a true equilibrium (e.g. it returns after annealing to the isotopic phase) and its origin is associated with the quenched orientational disorder introduced by the network crosslinks.  The resulting structure is analogous to the correlated spin glass with the quadrupolar local symmetry of the order parameter \cite{Fridrikh1997,zhou_2021}. Experimentally, the characteristic size of such `domains' (i.e. the regions of correlated director orientation) is on the order of 1-2 $\mu$m \cite{Clarke1998}, consistent with the strongly scattering (white) appearance of polydomain nematic elastomers. 

Applying an external uniaxial stress to such a polydomain microstructure results in the polydomain-monodomain transition characterized by a stress plateau and the transition from a strongly scattering to a clear transparent appearance \cite{Clarke1998,Fridrikh1999,urayama_2009,Biggins2009,Biggins2012}. 
The imposition of a biaxial stretch on a sheet reveals an `in-plane liquid-like behavior' where, after an initial elastic response, the (true) stresses in the two axes are equal even when the two imposed stretches are different up to a limit \cite{tokumoto_2021}.  Furthermore, the value of the stress depends only on the areal stretch (the product of the two stretches) independent of the ratio.  The uniaxial stress response can be shown to be a particular manifestation of this in-plane liquid like behavior. The mechanism underlying this unusual behavior is revealed by light scattering to be domain-reorientation (from an initially equidistributed to a more anisotropic distribution depending on the imposed stretch) \cite{tokumoto_2021}. 

In addition to the inherent interest, domain reorientation-induced elastic softness manifests itself in increased impact resistance, vibration attenuation, enhanced surface adhesion, and other phenomena that are of interest to various applications \cite{saed_2021,Hiro2021}. 
Here we obtain further insight into the remarkable elastic softness and the associated domain reorientation of polydomain nematic elastomers by studying the classical Hertz indentation problem.  This is a unique mechanical test where the force-displacement relation is nonlinear (with exponent 3/2) even when the material is linear because the volume of material being probed increases with indentation depth. We expect, and indeed find that the 'soft' anelasticity of nematic elastomers makes the Hertz test response dramatically different.

%\section{Experiment}
\vspace{0.15cm}
\noindent \textbf{Experiment.} \ We work with the most common main-chain nematic elastomers based on the thiol-acrylate click chemistry and point-like flexible crosslinkers. Since this class of materials was first introduced in 2015, it has become the system of choice due to its robust and reproducible synthesis and properties \cite{Mohand2015, Mohand2017}. The details of the material are described in those publications in sufficient detail. One important matter we had to address is the `genesis' of the elastomer network.
It is important to have the network crosslinked in the isotropic phase, so it enters into the nematic phase with a set of quenched random crosslinks but no additional texture constraints. It would be ideal to achieve this by having the network swollen by a solvent during crosslinking and then de-swell the crosslinked solid, avoiding chain entanglement. However, we were unable to achieve this because we require a large bulky sample that is hard to de-swell.  So we crosslink the polymer in the isotropic melt, and live with a possibly entangled network that provides additional stiffness.

\begin{figure}
 \centering
\includegraphics[width=4in]{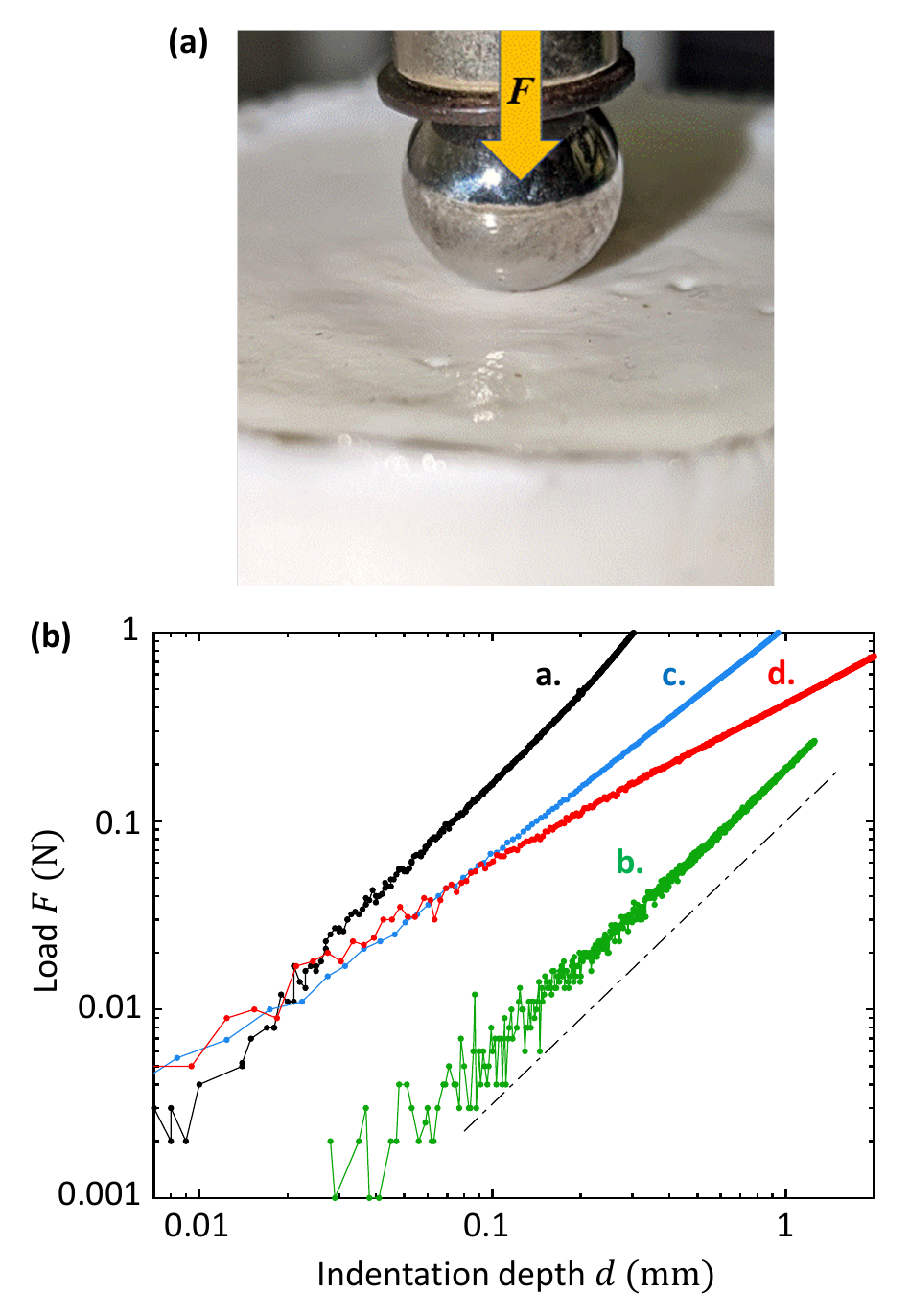}
\caption{(a) The illustration of our Hertz indentation test, with a solid sphere compressed into a flat thick layer of LCE, measuring the load and the vertical displacement. (b) A summary of experimental results, using the log-log scale to compare the ideal Hertz case for two isotropic elastomers (curves \textbf{a.} and \textbf{b.}) accurately following the slope $F \propto d^{3/2}$, and two LCE materials (curves \textbf{c.}, scaling exponent 1.26, and \textbf{d.}, scaling exponent 0.82).} 
\label{fig:Hertz}
\end{figure}

Figure~\ref{fig:Hertz} summarizes the results of our tests where a rigid spherical indenter (of radius $R$=5 mm) is driven into the thick elastomer disk (at a low speed $V$=0.05 mm/min, modeling equilibrium). 
Fig.~\ref{fig:Hertz}(a) illustrates the setup, and  Fig.~\ref{fig:Hertz}(b) the measured load $F$ vs. depth of indentation $d$ relation (on a log-log scale to identify the characteristic scaling). Two samples are isotropic elastomers -- a 10\%-crosslinked silicone rubber (curve \textbf{a.}) and a very weak polyacrylamide gel (curve \textbf{b.}), and both show the classical Hertz scaling \cite{timoshenko,Landau} $F = \frac{16}{9}E R^{1/2} d^{3/2}$, where $E$ is the Young modulus of the material.  The Young modulus are determined to be 1.3~MPa for \textbf{a.} and 0.47~kPa for \textbf{b.}, but the $d^{3/2}$ scaling is the same. 

The other two curves in Fig. \ref{fig:Hertz}(b) are obtained from the isotropic-genesis polydomain thiol-acrylate main-chain LCE pads (thickness 17 mm): a 'typical' 10\%-crosslinked LCE (curve \textbf{c.}), and  a weakly crosslinked LCE (curve \textbf{d.}).  It is clear that the scaling exponent of the $F \propto d^{x}$ relationship differs from the classical Hertz exponent of 3/2, and moreover material dependent. 
This is surprising since the exponent must be non-dimensional, meaning that material parameters (in units of MPa) must be reduced by another internal  parameter with dimensionality of energy density.
Further surprising, the exponents are smaller than 3/2: a softer mechanical behavior should lead to a higher, not lower, exponent at small indentation depths. 

Finally, the same indentation test carried out on LCE pads at a high temperature $T>T_\mathrm{ni}$ returns a perfectly Hertzian scaling of $F \propto d^{3/2}$ (not shown).  Thus, the unusual exponents are a manifestation of the nematic domains and their evolution.

%\section{Fitting model}
\vspace{0.15cm}
 \noindent \textbf{Theory.} \ We apply a coarse-grained theoretical model \cite{Lee2023}, which we briefly recall.  This model introduces two scalar state variables $\Lambda$ and $\Delta$ that describe the spontaneous deformation associated with the local domain pattern.  These are closely related with local polydomain order parameters: $\Lambda$ with the degree of orientation $S$, and $\Delta$ with $S+X$ where $X$ is the degree of biaxial orientation. These state variables describe the spontaneous change in material metric (the Cauchy-Green stretch due to domains) $\bm{G} = \bm{P} \ \text{diag}(\Lambda^2,\Delta^2/\Lambda^2,1/\Delta^2) \bm{P}^T$ where $\bm{P}$ is a rotation matrix, and $\Lambda$ and $\Delta$ can take values in the region $\{( \Delta \le r^{1/6}, \Delta \le \Lambda^2, \Delta \ge \sqrt{\Lambda}\}$ where $r$ is the chain anisotropy parameter (related to the degree of nematic order $Q$).  A monodomain has $\Lambda=r^{1/3}$ and $\Delta=r^{1/6}$ so that $\bm{G}$ is the step-length tensor $\bm \ell$ of the neo-classical theory \cite{WTbook}, and an isotropic polydomain state where the nematic directors are equidistributed has $\Lambda=\Delta=1$ so that $\bm{G}$ is identity.  The biaxial polydomain state where all the nematic directors are confined to a plane but equidistributed in the plane has $\Lambda=r^{1/12}$ and $\Delta=r^{1/6}$ so that $\bm{G}= \bm{P} \ \text{diag} (r^{1/12}, r^{1/12}, r^{-1/6})\bm{P}^T$.
 
 The model postulates a coarse-grained free energy $W=W_e+W_r$ where $W_e = \frac{1}{2}\mu [\text{tr} (\bm{F}^T \bm{G}^{-1} \bm{F}) - 3]$ is the entropic energy in the polymer chains for a deformation gradient $\bm{F}$ relative to an isotropic reference state, with $\mu$ the rubber modulus, and $W_r = C(\Delta-1)/(r^{1/6}- c \Delta )^k$ is the energy of domain patterns required to overcome fluctuations. For a monodomain LCE this reduces to the semi-soft theory discussed in \cite{WTbook} in some detail. The deformation is determined by the equation of mechanical equilibrium while the state variables evolve according to overdamped dynamics
 $\alpha_\Lambda \dot \Lambda = -\partial W/\partial \Lambda, \ \alpha_\Delta \dot \Delta = - \partial W/\partial \Delta$.  The model has been validated against experiments and verifiably implemented as a UMAT in the finite element package ABAQUS \cite{Lee2023}.

\begin{figure}
\centering
\includegraphics[width=5in]{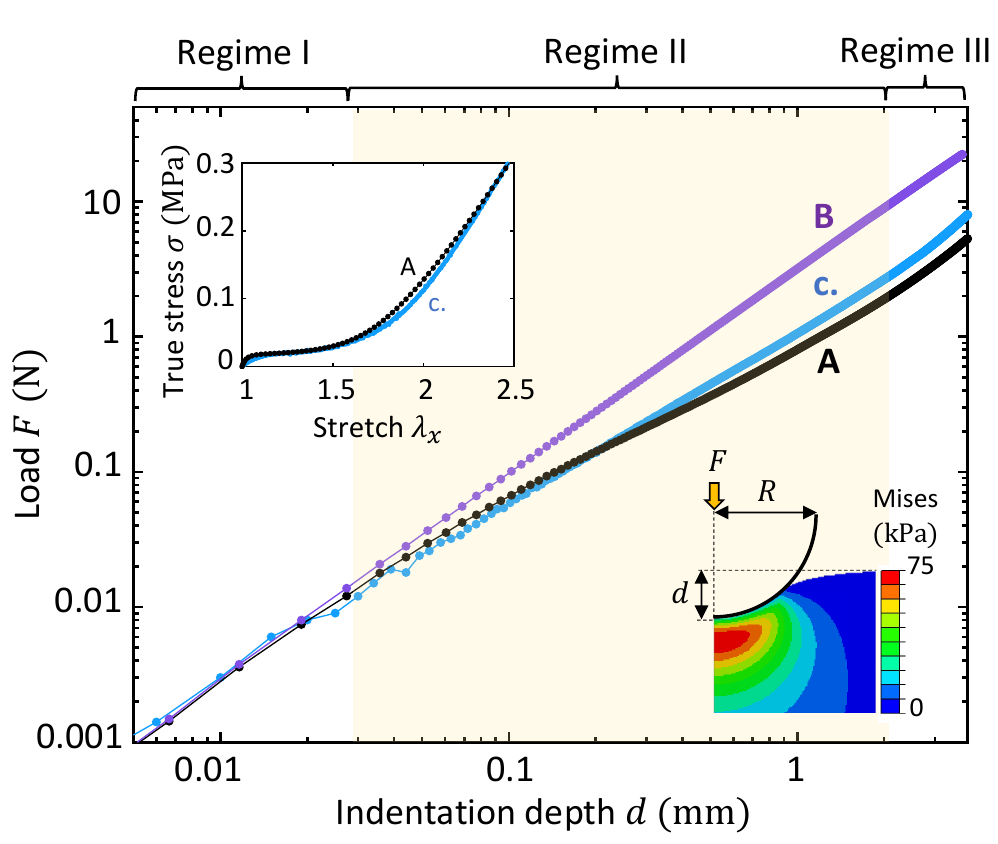}
\caption{The computed load vs. displacement relationship from the theoretical model ($\textbf{A}$, black) and comparison with experiment ($\textbf{c.}$, blue).  The model parameters for the model are fit to an independent uniaxial tensile test (top-left inset) on 10\%-crosslinked polydomain LCE and listed in Table \ref{tab:param} as material $\textbf{A}$.  The calculated results show three regimes: an initial Regime I with classical exponent 3/2, an intermediate softening Regime II with exponent 1.18, and a final Regime III with classical exponent 3/2.  The result for the corresponding isotropic model, same parameters except $r=1$ ($\textbf{B}$, purple) shows the classical exponent of 3/2. The bottom-left inset shows the computational setup and a snapshot of the state of stress around the contact region. 
\label{fig:theory}}
\end{figure}

\begin{table}
\centering
\caption{Parameters for cases A, B, C, and D. \label{tab:param}}
\begin{tabular}{l cccc} 
\hline % inserts single horizontal line
Parameter & \textbf{A} & \textbf{B} & \textbf{C} & \textbf{D}\\
\hline % inserts single horizontal line
Shear modulus $\mu$(MPa)  & 0.26 & 0.26 & 0.26 & 0.13\\
%Poisson ratio $\nu$  & 0.5 & 0.5 & 0.5 & 0.5\\
Anisotropy parameter $r$ & 6 & 1 & 6 & 6\\
Hardening coefficient $C$(kPa) \ \ \  & 0.6 & 0.6 & 0.6  & 0.1\\
Hardening coefficient $c$ & 0.95& 0.95  & 0.95 & 0.95 \\
Rate coefficient $\alpha_\Delta$(MPa.s) & 30 & 30 & 150 & 6\\
Rate coefficient $\alpha_\Lambda$(MPa.s) & 0.30 & 0.30 & 1.5 & 0.06\\ 
%Uniaxial strain rate $\dot{\epsilon}$(s$^{-1}$) & 0.0025 & 0.0125 & 0.0025\\
%Indentation speed $V$ (mm/min) \ \ \ & 0.05  & 0.05  & 0.25 & 0.05\\
\hline % inserts single horizontal line
\end{tabular}
\end{table}

We fit the model to the tensile stress-strain curve of the basic 10\%-crosslinked nematic elastomer, shown in the top-left inset of Fig. \ref{fig:theory}. These parameters are shown in Table \ref{tab:param} as material $\textbf{A}$. An axisymmetric model of a 17 mm-thick cylindrical LCE pad is created and discretized using four-node bilinear axisymmetric hybrid elements.  The lateral and top surfaces are traction-free while the bottom surface is fixed.   A 5 mm-radius indenter, modeled using axisymmetric rigid elements, is lowered into the LCE pad at $V=0.05$ mm/min as in experiment.  The contact is assumed to be friction-free.   The bottom-right inset of Fig. \ref{fig:theory} shows a snap-shot of the stress distribution at a cross-section in a region in the vicinity of the contact area.

The main Fig.~\ref{fig:theory} shows the computed force as a function of the indentation depth as the curve marked $\textbf{A}$.  We see three regimes: Regime I with exponent (slope in the log-log curve) 3/2 for very shallow indentation,  Regime II with an exponent 1.18 for intermediate indentation, and Regime III where the exponent returns towards 3/2 for quite deep indentation, when $d \sim R$.  The figure also shows the result of an analogous calculation with an elastic material with $r=1$ (parameters in Table \ref{tab:param}) in the curve marked $\textbf{B}$.  We see that the exponent is 3/2 for the elastic material as anticipated in the classical theory, thereby providing a verification of the method.  Finally, the figure reproduces the experimental observation as the curve marked $\textbf{c.}$: we see good agreement in Regimes II (the exponent in experiment is 1.23) and III. The agreement is significant since the model parameters were determined from an independent tensile test.
The experimental data is noisy in Regime I as forces are too small and surface roughness introduces artifacts.

\begin{figure}
 \centering
\includegraphics[width=5in]{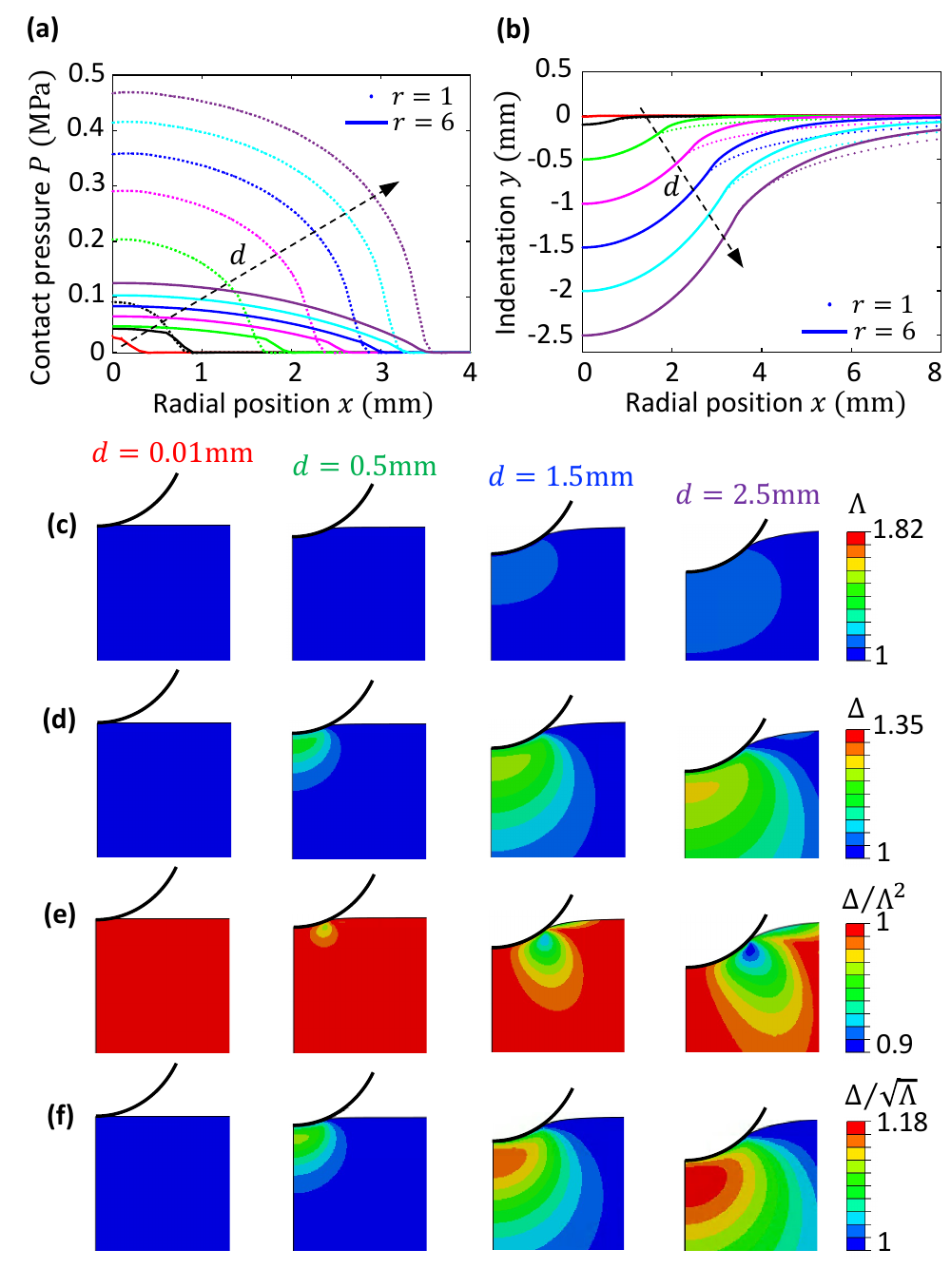}
\caption{Finite element simulation predictions of contact pressure (a) and surface topography (b) for two different substrates at various indentation depths $d$=\{0.01, 0.1, 0.5, 1, 1.5, 2, 2.5\} mm. Solid-line curves correspond to material $\textbf{A}$ with $r=6$, and the dotted-line curves correspond to material $\textbf{B}$, the isotropic version of material $\textbf{A}$ with $r=1$; both listed in Table I and shown in Fig. \ref{fig:theory}. (c-f) Distribution of the state variables $\Lambda$ (c), $\Delta$ (d), and ratios $\Delta/\Lambda^2$ (e) and $\Delta/\sqrt{\Lambda}$ (f) for material $\textbf{A}$ near contact region at a series of indentation depths.}
\label{fig:pressure}
\end{figure}

Figure~\ref{fig:pressure} provides details of the calculation that are not accessible to the experiment, but reveal the physics underlying the indentation response of the LCE, and the deviations from Hertzian behavior. Fig.~\ref{fig:pressure}(a) compares the distribution of pressure at the contact surface at various indentation depths for the LCE ($\textbf{A}$) and its isotropic version ($\textbf{B}$). The distribution of pressure for isotropic elastomer follows Hertz theory: the maximum pressure at the center of contact area $P_0\propto d^{1/2}$, and the pressure fall in a quadratic fashion away from the center \cite{timoshenko}. In contrast, the contact pressure profile for the LCE is very different: the peak is much smaller than the Hertzian prediction and the distribution more uniform. Consistent with Fig. \ref{fig:theory}, the contact pressure profiles of the LCE and elastic material agree in Regime I (at $d=0.01$mm)  reflecting the initial elastic response of the LCE.

Figure~\ref{fig:pressure}(b) shows the corresponding surface displacement profiles.  Consistent with Figs.~\ref{fig:theory}(b) and \ref{fig:pressure}(a), the surface topography of the LCE and elastic material agree in Regime I .  However, they deviate from one another in Regime II.  The contact radius for the LCE is larger than that for elastic material, and the deformation decays faster away from the indenter. 
The more uniform contact pressure distribution and its faster decay suggest that the semi-soft response of the LCE enables the deformation to be localized in the region below the indenter. In contrast, the deformation is much more delocalized in the elastic material.

Panels (c)-(f) of Fig.~\ref{fig:pressure} show the distribution of the state variables $\Lambda$ and $\Delta$, and the ratios $\Delta/\Lambda^2$ and $\Delta/\sqrt{\Lambda}$ at a series of indentation depths, and provide  insights into the nature of evolution of the domain pattern.  The color scale in Figs.~\ref{fig:pressure}(c,d) are chosen so that blue corresponds to the smallest value ($1$ for both) while red corresponds to the largest theoretical value ($r^{1/3} = 1.82$ for $\Lambda$, $r^{1/6} = 1.35$ for $\Delta$).  As the indenter is lowered into the LCE, the local biaxiality $\Delta$ evolves in the region below the indenter -- the region grows as does the peak value reaching the saturation value of ~1.3 (this is smaller than the theoretical peak value due to hardening).  In contrast, $\Lambda$ does not evolve too much.  In fact, we see from Fig. \ref{fig:pressure}(e,f) that $\Delta/\Lambda^2=1$ except in an annular region near the contact.  Therefore, the domain pattern evolves biaxially below the indenter till it saturates to a biaxial pattern at a certain depth of indentation, while it evolves towards a monodomain state in the annular region ($\Delta/\sqrt{\Lambda} \ =\ 1$).  As the indenter is pushed down, the soft behavior causes the region below the indenter to contract axially and expand laterally. This lateral expansion is accommodated by the soft shear response in the annular region thereby localizing the deformation (see Fig. S1 in Supplemental Material). 
We conclude that the domain evolution makes the spherical indenter behave like a cylindrical punch, resulting in a lower exponent in the intermediate Regime II. This effect is more pronounced when we increase the anisotropy ratio $r$ even resulting in a a bulge out of the free surface in the  vicinity of the contact circle (see Fig. S2 in Supplemental Material).

Taken together, the calculations reveal the physics of the three regions.  At very shallow penetration, corresponding to the pre-plateau regime of the tensile stress-strain curve, the material is elastic and therefore one observes a classical Hertz exponent of 3/2.  As penetration proceeds, the soft behavior kicks in as the domain pattern changes from triaxially equiaxed to biaxially equiaxed in the region below the indenter, and the exponent drops from 3/2 to a smaller number. There is a corresponding soft plateau on the tensile stress-strain curve. Finally, as the domains become fully biaxial under the indenter, the soft behavior saturates and the material become elastic again returning the response to the classical exponent of 3/2 (again, resembling the high-deformation linear regime in the tensile test).

\begin{figure}[t]
 \centering
\includegraphics[width=4in]{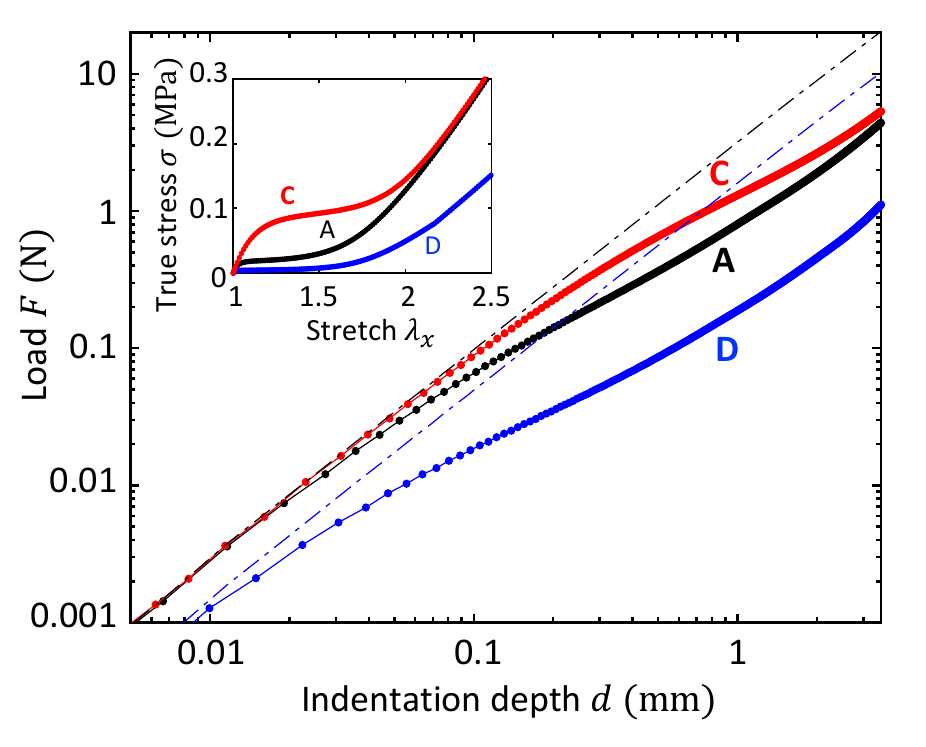}
\caption{The tensile stress-strain data (inset) and corresponding Hertz indentation tests using FE simulations show the non-universal scaling exponent of the $F \propto d^{x}$ for nematic polydomain LCEs. The black curves correspond to material $\textbf{A}$ shown in Fig. \ref{fig:theory}, the red curves present the tensile test and Hertz indentation results on a stiffer material $\textbf{C}$, and the blue curves corresponds to a softer material $\textbf{D}$ with parameters in Table I. The dashed lines show the reference $d^{3/2}$ scaling of isotropic elastomers.}
\label{fig:nonuniversality}
\end{figure}

 \vspace{0.15cm}
 \textbf{Universality}. We now examine if this three-region structure is universal, and further, whether the exponent in Regime II is universal.  It is difficult to explore Regime I and III experimentally due to artifacts.  However, experiments do reveal that the exponent in Regime II is not universal (Fig.~\ref{fig:Hertz}, curves \textbf{c.} and \textbf{d.}).  
So we examine this through simulation.  We consider three different sets of material parameters for the LCE: 
material $\textbf{A}$ as before, a stiffer material $\textbf{C}$, and a softer material $\textbf{D}$ with parameters in Table \ref{tab:param} and uniaxial stress-strain response in the inset of Fig.~\ref{fig:nonuniversality}.   Figure~\ref{fig:nonuniversality} compares the indentation response of the three materials.  It is clear that there is a universal structure of three regimes, though the indentation depth that separates these regimes shift left with softer material and right with the stiffer material.  However, the exponent in Regime II is not universal, and increases with stiffness.  This is consistent with the experimental observations.

%\section{Discussion}
\vspace{0.15cm}
\noindent \textbf{Discussion and conclusions.} 
In this paper, we study the indentation response in polydomain nematic elastomers through theory and experiment, and show how the highly  non-linear stress-strain response due to the local domain evolution causes a deviation from the classical Hertzian response.

It is common in the literature to attribute deviations from the classical Hertzian response due to adhesion, and indeed this is taken as a measure of the surface adhesion energy $\gamma_0$. For instance, the Johnson-Kendall-Roberts (JKR) theory \cite{JKR,Israelachvili2016} predicts the following $\gamma_0$-dependent correction in the isotropic material
\begin{equation}
d^{3/2} = \frac{9}{16 R^{1/2}E} \left( F + 6\pi R \gamma_0 + \sqrt{12 \pi R F \gamma_0 + (6\pi R \gamma_0)^2} \right)  \nonumber
\end{equation}
This predicts a lower load $F$ for any depth $d$ consistent to our observations; however, the JKR formula predicts a larger exponent in contradiction to our observations.  Furthermore, trying to find the adhesion energy from the measured pull-off force using the industry-standard condition of zero square root in the JKR formula, gives an unreasonably high $\gamma_0=25\,\mathrm{N/m}$ \cite{Hiro2021} (for comparison, the high-energy water-air interface has $\gamma_0=0.07\,\mathrm{N/m}$). All of this, along with our controls and theoretical analysis shows that adhesion is not the reason for our observations, and the JKR theory can not be used to find the surface energy $\gamma_0$ from indentation tests in LCEs.  

At the same time, LCE surfaces are characterized by a high ``stickiness''.  This is also consistent with the  much lower values of force $F(d)$ that we obtain in LCEs explain the high `stickiness' of their surface, because the physical response to contact load and its withdrawal present themselves as the enhanced adhesion.   Thus, our work reveals an alternate mechanism of enhanced adhesion, and a link between the Hertz contact mechanics in nematic LCEs and their pressure sensitive adhesion.

\vspace{0.2in}
\noindent \textbf{Author contributions.}
EMT conceived the probject and conducted the experiments with MOS.  AM conducted the simulations with assistance from KB.  All authors analyzed the results and wrote the paper.

\vspace{0.2in}
\noindent \textbf{Acknowledgements.}
This work has been funded by ERC H2020 (Advanced grant 758669), and as well as the US Office of Naval Research (MURI grant N00014-18-1-2624).

\newpage
\renewcommand\thefigure{S\arabic{figure} \Alph{section}}
\setcounter{figure}{0}
\renewcommand\thepage{S\arabic{page}}
\setcounter{page}{1}

\begin{center}
\text{Supplementary Material for:} \\
{ Softening of the Hertz indentation contact in nematic elastomers }\\
\text{\small Ameneh Maghsoodi, Mohand O. Saed, Eugene M. Terentjev, and Kaushik Bhattacharya}
\end{center}

\vspace{0.8cm}
%-------------------------------------------------------%
Figure S1 illustrate the strain components in the region below indentation of material {\textbf A} using left Cauchy-Green deformation tensor $\textbf C=\textbf F^T \textbf F$. As shown in Fig. S1, the region below the indenter expands biaxially in plane $xz$ with no shear deformation ($C_{xz}=0)$. This concludes that as the indenter is pushed down, the soft behavior of LCE substrate causes the domain pattern below the indenter to evolve biaxially. This effect is more pronounced when considering a very soft LCE substrate, as shown in Fig. S2. Figure S2 illustrates the simulation predictions for the material $\textbf E$ with a higher anisotropy $r=10$. Remarkably, as illustrated in Fig. S2(c), the soft behaviour of LCE results in a the bulging out of the free surface in the immediate vicinity of the contact circle. In other words, the indenter behaves more like a cylindrical punch than a spherical indenter. 

%--------------------------------------------
\begin{table} [h]
\caption{Material parameters. \label{tab:param}}
\centering
\begin{tabular}{l ccc} 
\hline % inserts single horizontal line
Parameter & \textbf{A}  & \textbf{E} & \textbf{H} \\
\hline % inserts single horizontal line
Shear modulus $\mu$(MPa) & 0.26 & 0.26 & 0.26 \\
Anisotropy parameter $r$  & 6 & 10 & 1\\
Hardening coefficient $C$(kPa) \ \ \ & 0.6 & 0.6 & 0.6 \\
Hardening coefficient $c$ & 0.95 & 0.95 & 0.95   \\
Rate coefficient $\alpha_\Delta$(MPa.s) & 30 & 1 & 1 \\
Rate coefficient $\alpha_\Lambda$(MPa.s) & 0.30 & 0.01 & 0.01\\ 
\hline % inserts single horizontal line
\end{tabular}
\end{table}

%-----------------------
\begin{figure}[h]
\begin{center}
  \includegraphics[width=\textwidth]{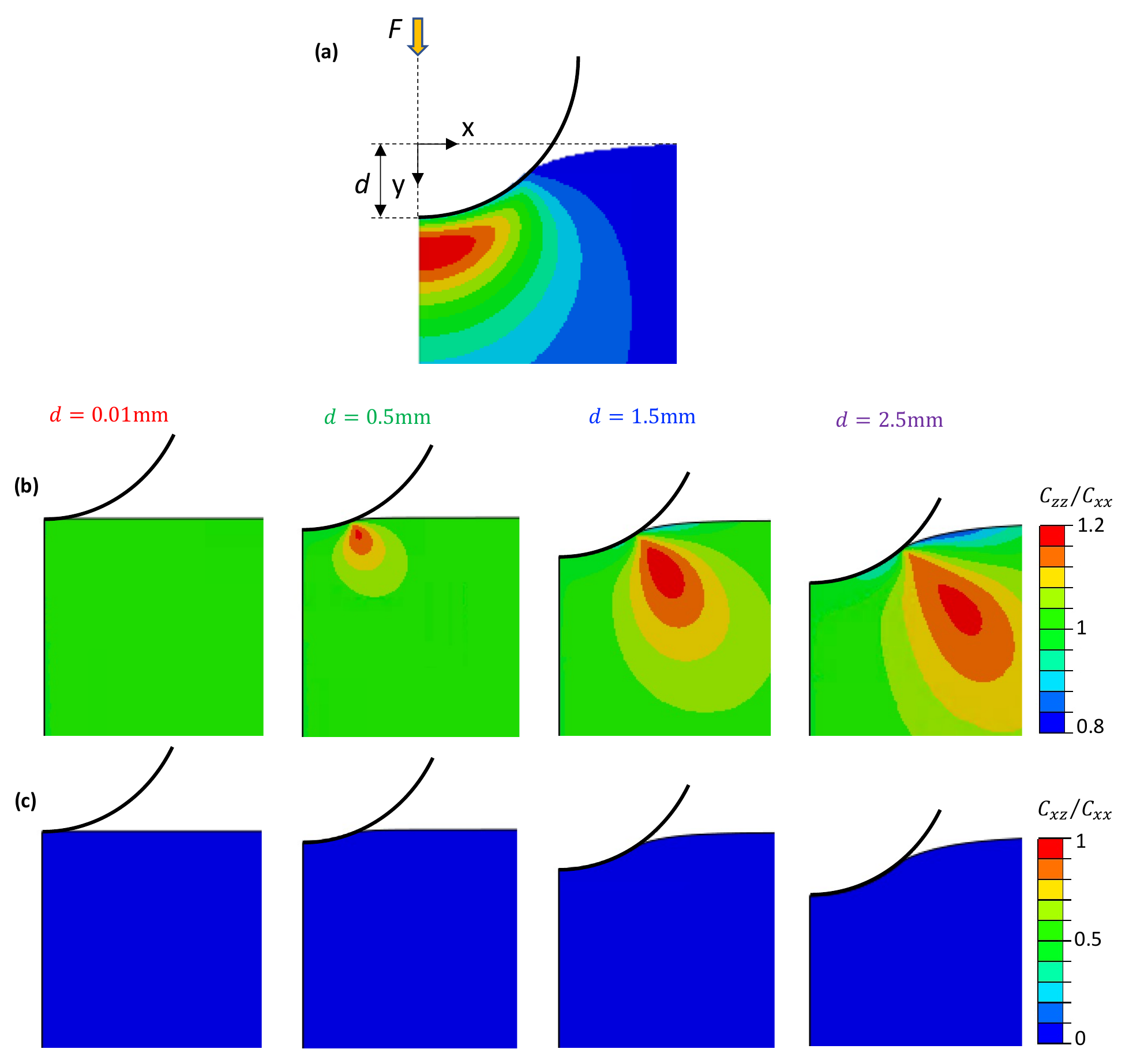}
  \caption{(a) Schematics of indentation problem. (b-c) The ratio of left Cauchy-Green components in plane $xz$ at various indentation depths $d$=\{0.01, 0.5, 1.5, 2.5\} mm for material {\textbf A} listed in Table 1 .}
  \label{FIG.S1}
\end{center}
\end{figure}
%------------------------

%-----------------------
\begin{figure}[h]
\begin{center}
  \includegraphics[width=\textwidth]{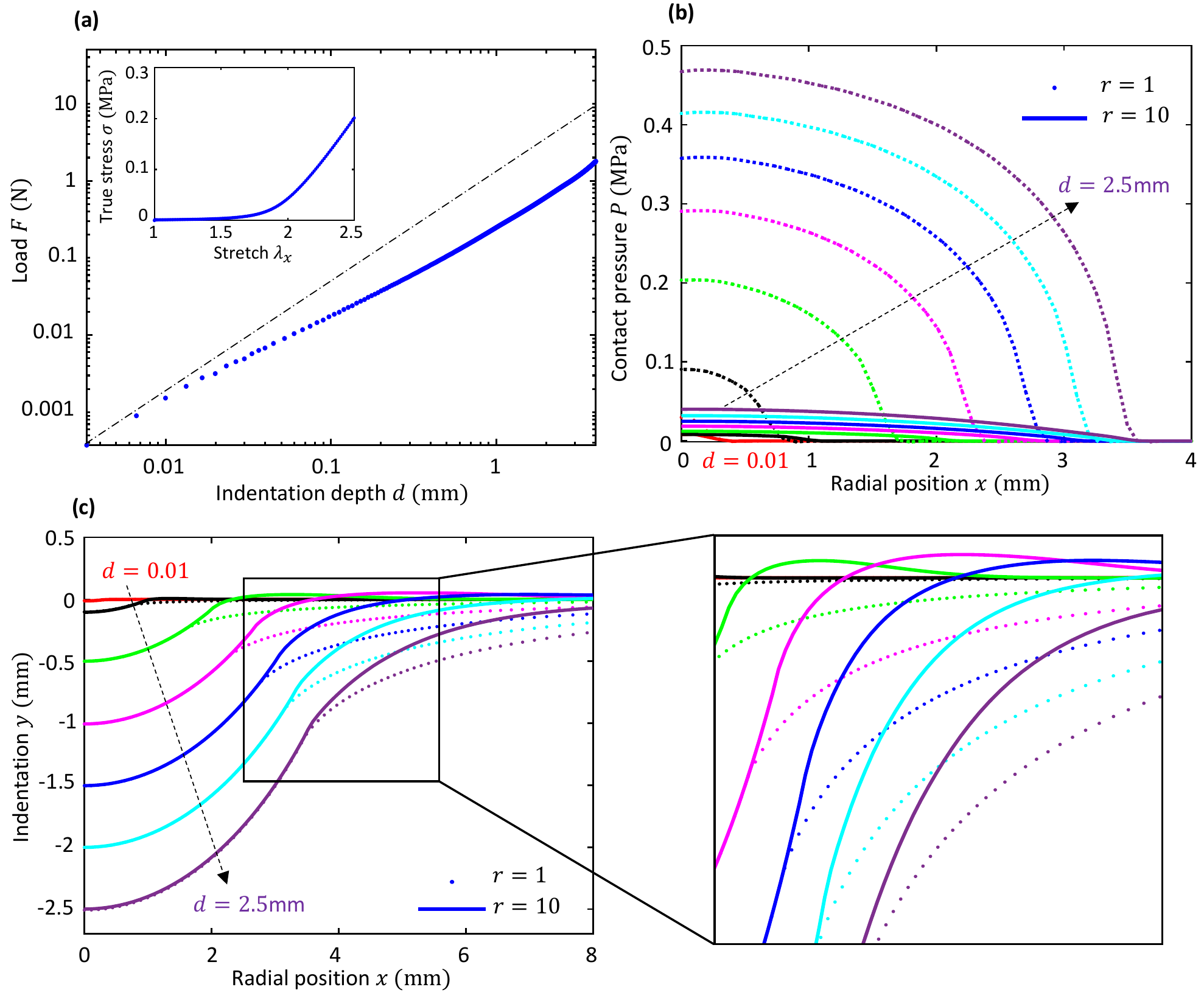}
  \caption{Finite element simulation predictions for LCE material \textbf{E} ($r=10$) and its isotropic version, material \textbf{H} ($r=1$), with parameters listed in Table 1. (a) The tensile stress-strain data (inset) and corresponding Hertz indentation tests for material \textbf{E}. Comparison of contact pressure (b), and surface topography (c) for two substrates at various indentation depths $d$=\{0.01, 0.1, 0.5, 1, 1.5, 2, 2.5\} mm. Solid-line curves correspond to the LCE material \textbf{E} and the dotted-line curves correspond to isotropic material \textbf{H}. The inset in (c) illustrate the bulging out of the free surface near contact area.}
  \label{FIG.S2}
\end{center}
\end{figure}
%------------------------

 \end{document}